\documentclass[11pt]{article}
\usepackage{amsmath,amsthm,latexsym,amssymb,amsfonts,parskip,epsfig}
\usepackage[a4paper,top=2.5cm,bottom=3.5cm]{geometry}
\allowdisplaybreaks
\numberwithin{equation}{section}

\newcommand{\N}{\mathbb{N}}                  

\newcommand{\expec}[1]{\left\langle #1 \right\rangle}

\newcommand{\at}{\text{atom}}
\newcommand{\lp}{l_\text{P}}
\newcommand{\dof}{\text{dof}\,}
\newcommand{\eff}{\text{eff}\,}
\newcommand{\lqg}{\text{lqg}\,}

\begin{document}
\title{Newton's constant from a minimal length: Additional models}
\author{Hanno Sahlmann\footnote{sahlmann@apctp.org}\\
{\small Institute for Theoretical Physics,}\\
{\small Karlsruhe Institute of Technology, Karlsruhe (Germany)}\\
{\small and}\\
{\small Asia Pacific Center for Theoretical Physics, Pohang (Korea)}}

\date{{\small PACS No.\ 04.60.-m, 04.60.Pp, 04.20.Cv}\\ 
{\small Preprint KA-TP-31-2010, APCTP Pre2010-006}}

\maketitle

\begin{abstract}
We follow arguments of Verlinde \cite{Verlinde:2010hp} and Klinkhamer \cite{Klinkhamer:2010qa}, and construct two models of the microscopic theory of a holographic screen that allow for the thermodynamical derivation of Newton's law, with Newton's constant expressed in terms of a minimal length scale $l$ contained in the area spectrum of the microscopic theory. One of the models is loosely related to the quantum structure of surfaces and isolated horizons in loop quantum gravity. Our investigation shows that the conclusions reached by Klinkhamer \cite{Klinkhamer:2010qa} regarding the new length scale $l$ seem to be generic in all their qualitative aspects.  
\end{abstract}
\section{Introduction}
The present work stands in the context of the recent attempt by Verlinde \cite{Verlinde:2010hp} to explain gravitational attraction as a thermodynamic effect. Introducing a fundamental length scale $l$ and using thermodynamic reasoning, Klinkhamer \cite{Klinkhamer:2010qa} then \emph{derives} the gravitational acceleration caused by a mass $M$, with the gravitational constant $G$ expressed in terms of the fundamental length scale $l$ as
\begin{equation}
G=f \frac{c^3l^2}{\hbar}.
\label{eq:newtong}
\end{equation}
The symbol $f$ in this formula stands for a numerical constant that is determined by the microscopic physics of the holographic screen replacing the mass $M$. To achieve this result, one expresses the rest energy of the mass causing the acceleration in terms of the energies stored in the degrees of freedom of the screen, 
\begin{equation}
Mc^2=\frac{1}{2}N_\dof k_B T,
\label{eq:thermo}
\end{equation}
with $T$ the Unruh temperature corresponding to the acceleration. For the derivation, a specific relationship between the number of degrees of freedom\footnote{Note however that in the models considered here and in \cite{Klinkhamer:2010qa}, $N_\dof$ is not an integer -- see also the discussion in section \ref{se:disc} on this point.} $N_\dof$, and the area $A$ of the screen, has to be satisfied, 
\begin{equation}
\frac{A}{fl^2}=N_\dof.
\label{eq:new}
\end{equation}
For the details of the derivation we refer to \cite{Klinkhamer:2010qa}. The equations \eqref{eq:thermo} and \eqref{eq:new} are new in the context of gravitational physics, and they have interesting consequences. Besides the derivation of Newton's law, there seems to be a relationship to the first law of black hole thermodynamics%
\footnote{Indeed, it is possible  to derive some form of first law of black hole thermodynamics from \eqref{eq:new} and \eqref{eq:thermo}, 
\begin{equation*}
\delta E = \frac{1}{2} kT \delta N_\dof =\frac{1}{2} kT\frac{1}{\lp^2}\delta A=
2T\delta S. 
\end{equation*}
The pre-factor is certainly wrong, so there is more to be understood.},
and \eqref{eq:new} constrains the microscopic structure of the screen. 
Indeed, in \cite{Klinkhamer:2010qa} a model was investigated in which the screen consisted of `atoms' of area $l^2$, and with an associated internal state space of dimension $d$. It turns out that $d$ is constrained by \eqref{eq:new} to one of two values.  

In the present short article we investigate the ramification of \eqref{eq:new}, by studying two new models of the microscopic structure of the screen. We find that \eqref{eq:new} restricts the microscopic theory even more strongly in our models, as it \emph{uniquely} determines the size of the internal state space. Our investigation also shows that the conclusions reached in \cite{Klinkhamer:2010qa} regarding the new length scale $l$ seem to be generic in all their qualitative aspects. Additionally, the second model we study is loosely related to the quantum structure of surfaces \cite{Ashtekar:1996eg} and isolated horizons \cite{Ashtekar:2000eq,Engle:2010kt} in loop quantum gravity . Our calculations thus also show that condition \eqref{eq:new} can be applied in that context. We will discuss our findings more thoroughly in section \ref{se:disc}. In the following section we will specify the models, and determine their free parameters. 
\section{Models and results}
In \cite{Klinkhamer:2010qa}, a model for the micro-physics of the holographic screen was considered. In it, the screen was made up of `atoms' or `tiles' of area $A_\at=l^2$. Each of them had a $d$-dimensional space of internal states. It turned out that in this model, equation \eqref{eq:new}, and consistency with the Bekenstein-Hawking formula for the entropy of the screen in the case that it lies where the horizon of a black hole would be, determined the parameters of the model up to a twofold degeneracy, 
\begin{equation}
(d,l^2)\approx\begin{cases} 
(8.6132,\, 2.2498\times 10^{-69}\text{m}^2)  \\(1.4296,\, 3.7343\times 10^{-70}\text{m}^2) 
\end{cases}.
\end{equation}
The feature that distinguishes the two models that we consider from the one in \cite{Klinkhamer:2010qa} is the following: In \cite{Klinkhamer:2010qa}, there is only one type of atom that makes up the holographic screen. In our models, we will have different species of such atoms. Each  species is labeled by an integer $s$. It has a distinct area $A_\at(s)$ and an internal Hilbert space of dimension $d_\at(s)$. We take the atoms to be distinguishable. For the area spectrum we choose 
\begin{equation}
A_\at(s)=l^2s.
\label{eq_areaspec}
\end{equation}
The two models that we consider differ in the internal structure of the atoms. In the simpler model, each carries an internal Hilbert space of the same size, irrespective of the label $s$, 
\begin{equation}
d_\at(s)=d_1=\text{const.} 
\label{eq_mod1}
\end{equation}
In the second model, the internal structure of different atom species is different,  
\begin{equation}
d_\at(s)=s+d_2,
\label{eq_mod2}
\end{equation}
where $d_2$ is a constant. 

In either model, we assume that the screen has an area
\begin{equation}
A=l^2 N \qquad \text{with } N\in \N.
\end{equation}
As is implicit in \eqref{eq:newtong}, $f$ is given by the ratio
\begin{equation*}
f=\frac{\lp^2}{l^2}, 
\end{equation*}
where $\lp$ is the Planck length $\lp=\sqrt{G\hbar/c^3}$. 
\subsection*{Model 1}
We now proceed to calculating $l$ and $d$ (or equivalently, $l$ and $f$) in the first model. The number of ordered partitions of $N$ into $n$ parts is given by 
\begin{equation*}
p(N,n) = \binom{N-1}{n-1}. 
\end{equation*}
Hence the number of microstates of the screen is
\begin{equation}
\sum_{n=1}^Np(N,n)d_1^n=d_1(d_1+1)^{N-1}.
\end{equation}
Asking for the Bekenstein-Hawking relation $S/k_B=1/4\, N/f$ then gives  
\begin{equation}
\ln(d_1+1)=\frac{1}{4} f^{-1}.
\label{eq_bekenstein1}
\end{equation}
Now we come to the new relation \eqref{eq:new} from \cite{Klinkhamer:2010qa}. The new feature in our model is that it does not have a fixed number of degrees of freedom. Rather, the number of atoms changes with the microstate. 
What we will do in this situation is to ask for 
\begin{equation}
\frac{A}{\lp^2}=\expec{N_\dof},
\label{eq:new2}
\end{equation}
where $\expec{N_\dof}$ is the average of $N_\dof$ with equal weight for all microstates. This seems to be a reasonable way to generalize \eqref{eq:new} for a varying number of degrees of freedom, but it will ultimately have to be \emph{derived} from the statistical mechanics underlying \eqref{eq:thermo}.
A short calculation reveals that 
\begin{equation*}
\sum_{\text{microstates}}N_\at=d_1(d_1+1)^{N-1}+d_1^2(N-1)(d_1+1)^{N-2},
\end{equation*}
whence 
\begin{equation}
\expec{N_\at}=1+\frac{d_1(N-1)}{d_1+1},
\label{eq_naver1}
\end{equation}
and with $\expec{N_\dof}=d_1\expec{N_\at}$ we obtain the transcendental equation 
\begin{equation}
\ln(d_1+1)=\frac{1}{4}\frac{d_1^2}{d_1+1}
\label{eq_trans1}
\end{equation}
for the number of internal degrees of freedom of the atoms. This equation has a spurious solution $d_1=0$ (giving $f=\infty$, $l=0$), and the physical solution 
\begin{equation}
d_1\approx 10.7811. 
\end{equation}
The physical solution then determines
\begin{equation}
f\approx 0.1014, \qquad l^2\approx 2.5773\times 10^{-69} \text{m}^2. 
\end{equation}
\subsection*{Model 2}
In this model, microstates can be labeled partially by sequences $(n_s)_{s=1}^{\infty}$, where $n_s$ is the number of atoms of species $s$. Thus $N_\at=\sum_s n_s$. Since the atoms are distinguishable and have internal states, there is a degeneracy associated to such a sequence,
\begin{equation}
d_{(n_s)}=\frac{\left(\sum_s n_s\right)!}{\prod_s n_s!}\prod_s(s+d_2)^{n_s}.
\label{eq:degen}
\end{equation}
The degeneracy is maximized by a certain assignment $s\mapsto n_s$. To do the calculations for this model, we use an approximation devised by Ghosh and Mitra \cite{Ghosh:2006ph} that has turned out to be very accurate and useful in loop quantum gravity. We will only consider those configurations $(n_s)$ which maximize this degeneracy. These configurations can be determined by maximizing $d_{(n_s)}$ under the condition that the total area $\sum_s l^2 n_s s$ is held at the fixed value $A$. The result is most easily expressible in terms of the relative numbers of atoms,   
\begin{equation}
\widehat{n}_g:=\frac{n_g}{N_\at}=(g+d_2)e^{-\lambda l^2 g}, 
\label{eq:max}
\end{equation}
where $\lambda$ is a Lagrange multiplier related to the constraint on the area. We will determine $\lambda$ momentarily. To this end, we note that  
\begin{equation}
\sum_s\widehat{n}_s=1
\label{eq:norm}
\end{equation} 
gives the additional equation that can be used to determine $\lambda$. We calculate 
\begin{equation*}
\sum_s\widehat{n}_s=\frac{x(d_2+1-d_2x)}{(x-1)^2} 
\qquad \text{with } 
x=\exp(-\lambda l^2), 
\end{equation*}
but instead of solving \eqref{eq:norm} for $\lambda$, we solve for $d_2$ first, because it makes the calculation easier:
\begin{equation}
d_2=\frac{1-3x+x^2}{x(1-x)}.
\label{eq:d}
\end{equation}
Evaluating the degeneracy $d_{(n_s)}$ at the solution \eqref{eq:max}, and using Stirling's approximation, gives
\begin{equation*}
d_{(n_s)}=\frac{\sqrt{2\pi N_\at}}{\prod_s\sqrt{2\pi n_s}}e^{\lambda A}+ \text{lower orders in $N$}.
\end{equation*}
For the case that the screen coincides with a black hole horizon, we can read off $\lambda$,  
\begin{equation*}
\lambda=\frac{1}{4\lp^2}. 
\end{equation*}
We also note that thus $x=\exp(-1/(4f))$. We now come to the relation \eqref{eq:new}. Since we are just considering the maximally degenerate configurations of the horizon, the number of degrees of freedom is actually fixed, 
\begin{equation}
N_\dof=\sum_s n_s(s+d_2)=\frac{A}{l^2}+d_2N_\at=N+d_2N_\at.
\label{eq:ndof}
\end{equation}
We determine $N_\at$ from the relation
\begin{equation*}
A=l^2 N_\at \sum_{s}(s+d_2)x^s=l^2 N_\at \frac{x (d_2x-d_2-x-1)}{(x-1)^3}.
\end{equation*}
Solving for $N_\at$, and inserting into \eqref{eq:ndof}, we find 
\begin{equation*}
N_\dof=N\left( 1+d_2 \frac{(1-x)^3}{2x^2-2x+1}\right),
\end{equation*}
where we have also used \eqref{eq:norm}. Eliminating $d_2$ via \eqref{eq:d}, and recalling $f^{-1}=-4\ln x$, we finally obtain the equation 
\begin{equation}
-4\ln x=1+\frac{(x-1)^2(1-3x+x^2)}{x(2x^2-2x+1)}.
\end{equation}
As in the first model, this is a transcendental equation with two solutions, 
\begin{equation*}
x\approx 0.79091 \text{ and } x\approx 0.0840645,
\end{equation*}
of which the first one is spurious, as it leads to a negative $d_2\approx -4.51825$. The second solution gives 
\begin{equation}
d_2\approx 9.80385, \qquad f\approx 0.100962, \qquad l^2\approx 2.58737 \times 10^{-69} \text{m}^2. 
\end{equation}
\section{Discussion}
\label{se:disc}
We have studied two models of the microscopic structure of a holographic screen. In both cases, there are two solutions for the parameters of the models, but in each case one of them is clearly unphysical, leaving one with a \emph{unique} value for the parameters such that a derivation of Newton's law as in \cite{Klinkhamer:2010qa} can be carried out. 
It is remarkable, that the corresponding equations have exactly one physical solution. A priori, it is well possible that there are many, or there is none at all. 

Our investigation also shows that the conclusions reached in \cite{Klinkhamer:2010qa} regarding the new length scale $l$ (few possible values, of the order of, but not equal to $\lp$, etc.) seem to be generic. \cite{Klinkhamer:2010qa} also contained the hypothesis that, perhaps, the model considered there would apply to a large class of microscopic theories \emph{effectively}, thus giving the correct  values for a suitably defined effective length scale. While we have not found a way to do this \emph{exactly} for our models, it is not implausible that such a definition of an effective length scale exists.  
To give at least one example, in our first model, one could set 
$l_\eff:=A/\expec{N_\at}$. Then, for large $A$, one has 
$l^2_\eff \approx l^2(d_1+1)/d_1\approx 2.8163 \times 10^{-69}\text{m}^2$.  But many other reasonable definitions for the effective length scale exist, and one of them could indeed reproduce the results of \cite{Klinkhamer:2010qa} exactly. 

We should remark that both models require a slight generalization of \eqref{eq:new} (in the first model: \eqref{eq:new2}; in the second model the problem did not arise due to the chosen approximation) due to the fact that the number of atoms is not fixed. While \eqref{eq:new2} seems to be  reasonable, it will ultimately have to be \emph{derived} from the statistical mechanics underlying \eqref{eq:thermo}. 

It is interesting to note that the situation we have considered is somewhat similar to that in loop quantum gravity. There, surfaces are endowed with area through intersections with one-dimensional excitations of the quantum gravitational field. These excitations can be labeled by integers $s$, and a single, simple excitation of type $s$, terminating on a surface, contributes an area quantum 
\begin{equation}
A(s)=2\pi\gamma \lp^2 \sqrt{s(s+2)}
\label{eq:lqg}
\end{equation}
and can have $d(s)=s+1$ internal states. 
The full area spectrum is slightly more complicated \cite{Ashtekar:1996eg} (and there are further differences in the case the surface is part of a black hole horizon). $\gamma$ is the so-called Barbero-Immirzi parameter, and is a priori free. 
This suggests that rather than viewing $G$, and perhaps $\gamma$, as fundamental in loop quantum gravity, one can take $\gamma$ and the length scale 
\begin{equation*}
l^2_{\lqg}=2\pi\gamma \lp^2.  
\end{equation*}
to be fundamental. Hence we are in a very similar situation as in the models that we investigated, with $\gamma$ playing the role of $1/(2\pi f)$. It should however be kept in mind that the theory for surfaces \cite{Ashtekar:1996eg} and isolated horizons \cite{Ashtekar:2000eq, Engle:2010kt} in loop quantum gravity is slightly more subtle than suggested by \eqref{eq:lqg}, with additional spectral values for the area, and a global constraint on states in the latter case.  

The interest in the effect of condition \eqref{eq:new} on the models considered here arose because of the thermodynamic argument in \cite{Klinkhamer:2010qa} leading to Newton's law, and to a formula for Newton's constant in terms of $l$. Such thermodynamic reasoning may at first glance seem irrelevant to loop quantum gravity, which attempts to quantize gravity in a fairly standard way (albeit with non-standard mathematical tools).  
But there is an unresolved tension already, as loop quantum gravity does find and count the number of microstates of a black hole as would be expected from the Bekenstein-Hawking relation, thus suggesting that the laws of black hole thermodynamics are indeed the laws thermodynamics. 
Thus thermodynamic reasoning may still be appropriate and useful in loop quantum gravity.

A last point worth mentioning is that in both models the dimension $d$ describing the internal spaces comes out non-integer. While this is perhaps acceptable for an effective description, it would not be for a fundamental theory. In this sense, the condition \eqref{eq:new} is very restrictive, and could be used to single out viable microscopic theories. For example, since loop quantum gravity does not possess a free parameter $d$, it would either fulfill \eqref{eq:new}, or not. While answering this question is feasible, it would be a rather complicated calculation, and it seems that one should further study the merit of \eqref{eq:thermo} and \eqref{eq:new} before applying it to loop quantum gravity, or any other theory of quantum gravity.    

\section*{Acknowledgments}
I thank Frans Klinkhamer for several important comments and discussions about \cite{Klinkhamer:2010qa} and the subject of the article, as well as for a careful reading of the draft. The research was partially supported by the Spanish MICINN project No.\ FIS2008-06078-C03-03.


\end{document}